# Plasmonics of magnetic and topological graphene-based nanostructures

*Dmitry A. Kuzmin\*, †, ‡, Igor V. Bychkov†, ‡, Vladimir G. Shavrov§, Vasily V. Temnov\* ¶,¶¶*

† Chelyabinsk State University, Department of Radio-Physics and Electronics, Br. Kashirinykh Street 129, 454001 Chelyabinsk, Russian Federation.

‡ South Ural State University (National Research University), 76 Lenin Prospekt, Chelyabinsk 454080, Russian Federation.

§ Kotelnikov Institute of Radio-engeneering and Electronics of RAS, 11/7 Mokhovaya Str., Moscow 125009, Russian Federation.

¶Institut des Molécules et Matériaux du Mans, CNRS UMR 6283, Université du Maine, 72085 Le Mans cedex, France

¶¶Groupe d'Etude de la Matiere Condensée (GEMaC), Université de Versailles-Saint Quentin en Yvelines, CNRS UMR 8635, Université Paris-Sacley, 45 avenue des Etats-Unis, 78035 Versailles Cedex, France

ABSTRACT. Graphene is a unique material to study fundamental limits of plasmonics. Apart from the ultimate single-layer thickness, its carrier concentration can be tuned by chemical doping or applying an electric field. In this manner the electrodynamic properties of graphene can be varied from highly conductive to dielectric. Graphene supports strongly confined, propagating surface plasmon-polaritons (SPPs) in a broad spectral range from terahertz to mid-infrared frequencies. It also possesses a strong magneto-optical response and thus provides complimentary architectures to conventional magneto-plasmonics based on magneto-optically active metals or dielectrics. Despite of a large number of review articles devoted to plasmonic properties and applications of graphene, little is known about graphene magneto-plasmonics and topological effects in graphene-based nanostructures, which represent the main subject of this review. We discuss several strategies to enhance plasmonic effects in topologically distinct closed surface landscapes, i.e. graphene nanotubes, cylindric nanocavities and toroidal nanostructures. A novel phenomenon of the strongly asymmetric SPP propagation on chiral meta-structures and fundamental relations between structural and plasmonic topological indices are reviewed.

I. Introduction

Graphene opens up wide prospects for numerous flatland photonic and plasmonic applications [1-7]. Graphene-based waveguides support localized electromagnetic SPP waves, both TE- and TM- polarized [8-16]. Their tight confinement and long propagation length allow for observing strong light-matter interactions in graphene-based structures [17]. Optical properties of graphene can be controlled by a number of external parameters such as the electrostatic bias, magnetic field or chemical doping.

For realizing any plasmonic device one should possess a tool for SPP manipulation. A conventional approach in noble metal-based active plasmonics relies on combining plasmonic and optically active materials [18-22]. For example, the coupling between magnetic and optical properties in magneto-optical materials leads to the optically induced magnetic fields through the inverse Faraday effect [23-26] or enhanced magneto-optical effects due to plasmonic excitations [27-34].

Spatial nanostructuring of graphene offers another unique playground known as plasmonic meta-surfaces. A periodic arrangement of densely packed sub-wavelength graphene stripes forms a meta-surface (see Figure 1A), which has a lot of promising plasmonic properties: it can generally support both elliptic (Fig. 1B) and hyperbolic SPPs (Fig. 1C) depending on various parameters such as the structure periodicity, its filling factor, graphene Fermi energy and frequency of exciting light [35-37]. In the hyperbolic regime, enhanced light-matter interactions are possible because of the strong spatial confinement of SPP electric field.

The intrinsic topological character (elliptic versus hyperbolic) of SPP dispersion on a meta-surface can be changed, for a given frequency, by tuning the optical properties through the highly anisotropic "epsilon-near-zero points", for example by using an external electric bias voltage [35-37]. This property is of great interest for the controlled guiding of SPPs.

Another way to merge topology with graphene plasmonics (and nanophotonics in general) is to create topologically distinct 3D-objects formed by 2D-meta-surfaces. Rolling up a meta-surface in a tube forms a meta-tube and bending the latter into the torus forms what we denote as a meta-torus (Fig. 1D). As we are going to show in this review article these geometrical transformations give rise to distinct topological indices, which largely determine the properties of plasmonic modes. The interplay of the intrinsic (elliptic versus hyperbolic) topology of SPPs propagating on a flat 2D-meta-surface and the geometrical 3D-topology of nanostructures can induce novel plasmonic effects: a giant azimuthal rotation of intensity distribution of particular SPP modes upon propagation, one-way propagation of SPPs, vanishing of the Fabry-Perot resonances in finite length meta-tubes, unidirectional circulating Mach-Zehnder-like resonances in a meta-torus, etc.

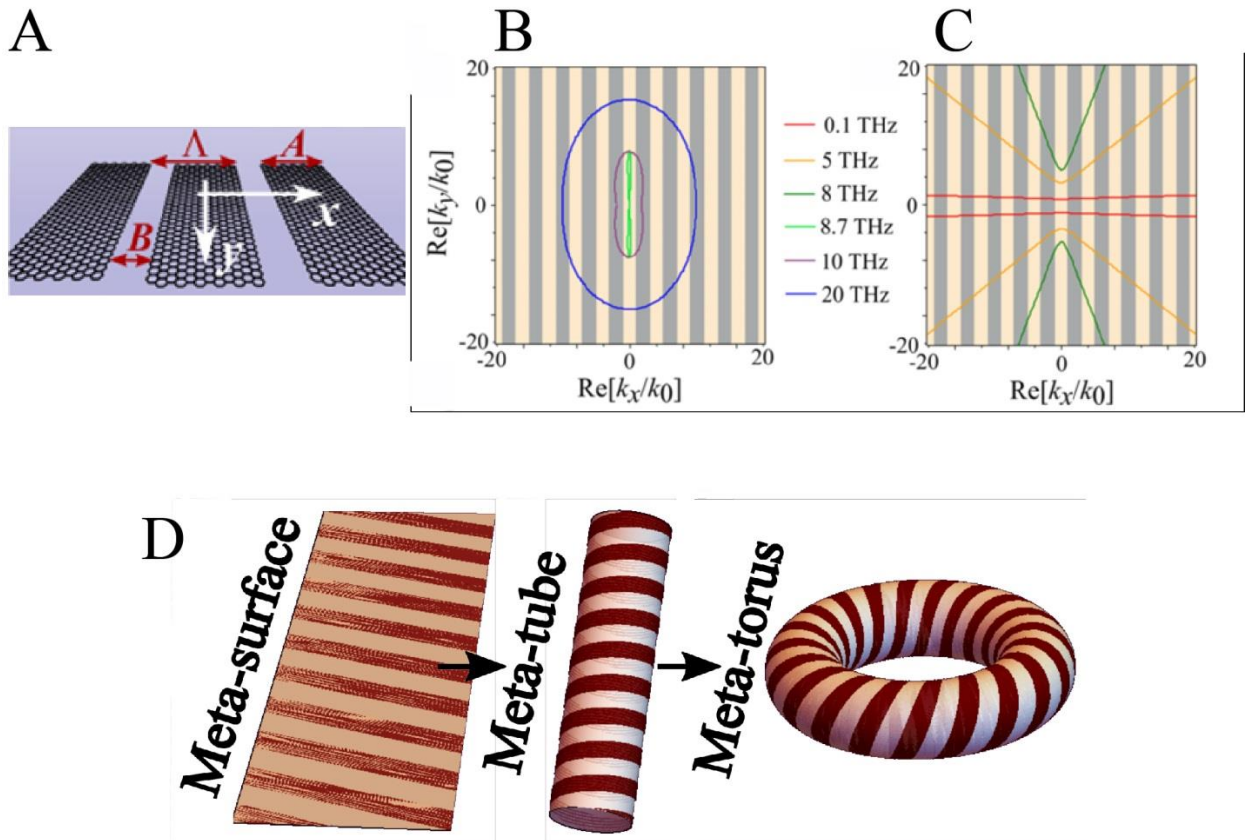

**Figure 1:** (**A**) An array of densely packed graphene stripes with sub-wavelength periodicity $\Lambda$ forms a metasurface which may support both elliptic (**B**) and hyperbolic SPPs (**C**). A rolled-up meta-surface forms a meta-tube and its donut-like shape is a meta-torus (**D**). In (**B**) and (**C**) $\Lambda =$ 50 nm, A = 45 nm, B = 5 nm.

The cylindrical geometry deserves a particular attention as it provides a common basis for plasmonics in chiral media [38-42] and non-reciprocal magneto-optics in waveguides. It is known that the parallel external magnetic field can rotate a spatially inhomogeneous intensity distribution (i.e. speckle-pattern) of light in the cross-section plane upon its propagation along an optical fiber [43-46]. Recently, we have demonstrated that also in graphene-coated optical fibers (with magneto-optical fiber material) one can control such a rotation by an external magnetic field. However, for an observable rotation it is necessary to have a fiber length of a few centimeters [47]. In line with the general trend of enhancing of the magneto-optical effects in magneto-plasmonic nanostructures it has been shown that the abovementioned rotation can be drastically enhanced in graphene-coated gyrotropic nanowires [48] because of the reduction of the wire radius results in a stronger confinement and increase of the wave vector of SPP modes. Magneto-optical effects in gyrotropic cylindrical structures have been investigated within the framework of the electromagnetic scattering problem [49, 50] and led to numerous applications in physics of optical devices [51-53]. It is worth mentioning recent plasmonic experiments performed on uniformly sized arrays of carbon nanotubes [54], which represent an ultimate size limit of nanoscaled graphene nanowires.

Abovementioned effects in the cylindrical geometry originate from the magnetic symmetry breaking for the azimuthal modes rotating in the opposite directions (i.e. with opposite signs of the azimuthal mode index $m$, or the so-called topological charge). An alternative way to introduce symmetry breaking is provided by the chirality. The chirality, which does not exist on flat surfaces, can be introduced in a twisted bilayer graphene [55, 56] or winding up a flat graphene-based meta-surface in a cylindrical meta-tube [57].

The intuitive picture of the aforementioned effects is illustrated in Figure 2. Chiral, azimuthal plasmonic modes propagating along the cylindrical structures are somewhat analogous to the nuts on the screws (Fig. 2A). Higher order plasmonic modes possess $2m$ nodes, giving the angular intensity distribution visually resembling the shape of common nuts (Fig. 2B). Whereas in the mechanical case the rotation direction of the nut is determined by the (left- or right-handed) thread on the screw, in plasmonics both rotation directions are generally possible giving rise to the propagating electromagnetic modes rotating clock (+$m$) or counterclockwise (-$m$). However, under appropriate conditions, in analogy to the mechanical nut-on-the-screw example, the chirality of the propagating plasmonic modes is dictated by the chirality of the meta-tube or by the direction of the magnetic field and modes with the opposite chirality cannot propagate.

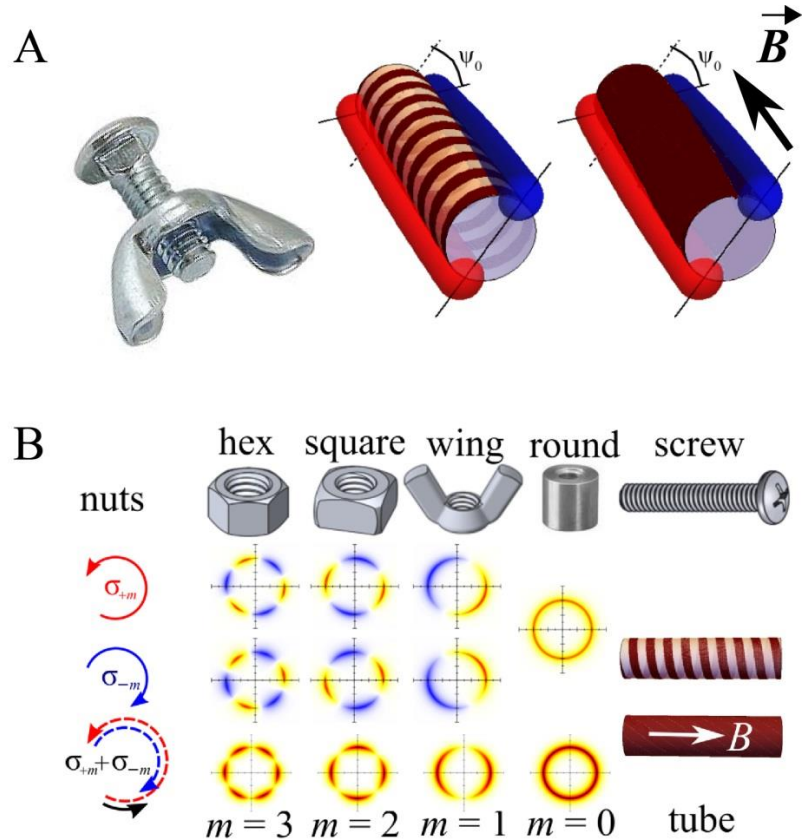

**Figure 2:** (**A**) Similar to a nut on a screw the spatially inhomogeneous azimuthal SPP intensity distribution (speckle pattern) rotates upon propagation along both chiral and magneto-optical

graphene-based nanowires. (**B**) Chiral $\sigma_{+m}$ and $\sigma_{-m}$ SPP modes characterized by opposite azimuthal indices $\pm m$ propagate along plasmonic nanowires with different k-vectors resulting in the asymmetry in the rotation of their electric field distribution upon propagation per unit length. More complex superpositions (speckle pattern) composed of several modes, like $\sigma_{+m}+\sigma_{-m}$, rotate upon propagation as well. The symmetry breaking between the chiral $\pm m$ SPP modes can be either due to the chirality in a graphene meta-tube or due to the magneto-optical activity of a dielectric core in a graphene-coated nanowire.

A linear superposition of azimuthal $\sigma_{+m}$ and $\sigma_{-m}$ modes creates a spatially inhomogeneous intensity distribution (speckle pattern) which is slowly rotating upon propagation along the wire due to the slightly different wave-vectors for $\sigma_{+m}$ and $\sigma_{-m}$ modes. Figure 2 illustrates that the $\sigma_{+m}+\sigma_{-m}$ superposition of SPP modes rotates similarly on a chiral meta-tube and a graphene-coated magneto-optical nanowire.

In the following we review the optical, magneto-optical and magneto-plasmonic properties of graphene, plasmonics of graphene-based meta-surfaces, magneto-plasmonics of graphene-based cylinders and novel plasmonic effects in topological graphene-based nanostructures. We believe that these results extend beyond the graphene plasmonics as they are qualitatively valid for arbitrary nanostructures formed by artificial 2D meta-surfaces supporting propagating SPP modes.

## II. Optical and magneto-optical properties of graphene

Dirac's character of quasiparticles in graphene leads to unusual dynamics of the electrons and holes. For example, the unconventional quantum Hall effect has been predicted theoretically [58-63] and observed experimentally [64-66]. Another feature is a finite effective cyclotron mass for the massless Dirac quasiparticles in the both electric and magnetic dc measurements which was found to vary as a square root of the number of carriers [64, 67-69]. From the experimental measurement of the graphene transmittance spectra [70-72], the dynamical conductivity was found to be frequency independent for the visible light $\sigma(\omega) = e^2/4\hbar$, which is in agreement with theoretical calculations [73-75].

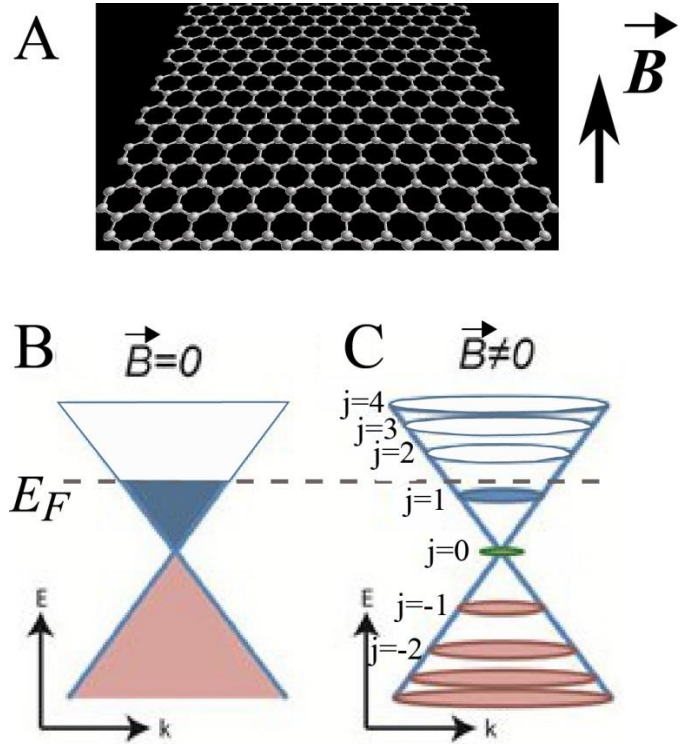

**Figure 3:** Graphene in an external perpendicular magnetic field (**A**) has a discrete set of non-equidistant electronic energy levels (Landau levels) in contrast to zero magnetic field (**B**).

From the point of view of theory, graphene is commonly described in terms of Dirac's gapless fermions. According to this picture, in graphene, there are two bands at the **K** hexagon vertices of the Brillouin zone without any gap between them, and the electron dispersion can be considered as linear in the wide wavevector region. Summing up the contributions of these points, i.e. integrating over the angle of the two-dimensional momentum vector **p** leads to the general quantum expression for the dynamic conductivity of graphene depending on both the frequency ω and the wave vector $k$ [76, 77] and meaning that both temporal and spatial dispersion should be taken into account. In the optical range, one can neglect the spatial dispersion of conductivity. This conductivity represents a sum of two contributions $\sigma(\omega) = \sigma_{intra}(\omega) + \sigma_{inter}(\omega)$. The first term $\sigma_{intra}(\omega)$ corresponding to the intraband electron-phonon scattering process, has a Drude-like behavior in the high frequency regime $\omega \gg \max(kv_F, \tau^{-1})$ [76-79], where $v_F$ is the carriers velocity (Fermy velocity) and $\tau$ is the carriers relaxation time

$$\sigma_{intra}(\omega) = \frac{2ie^2 k_B T \ln\left[2\cosh\left(\mu_{ch}/2k_B T\right)\right]}{\pi\hbar(\omega+i\tau^{-1})} \xrightarrow{\mu_{ch} \gg k_B T} \frac{ie^2 |\mu_{ch}|}{\pi\hbar(\omega+i\tau^{-1})} \quad (1)$$

while the second one, $\sigma_{inter}(\omega)$, corresponds to the direct interband electron transitions and plays the leading role around the absorption edge $\hbar\omega \approx 2\mu_{ch}$

$$\sigma_{inter}(\omega) = \frac{e^2}{4\hbar}\left\{\frac{1}{2} + \frac{1}{\pi}\arctan\left[\frac{\hbar\omega - 2\mu_{ch}}{2k_B T}\right] - \frac{i}{2\pi}\ln\frac{(\hbar\omega + 2\mu_{ch})^2}{(\hbar\omega - 2\mu_{ch})^2 + (2k_B T)^2}\right\}. \quad (2)$$

For room temperature $k_B T \sim 25$ meV, for THz, near-infrared and visible frequencies the photon energies are ~ 5 meV, 1 eV, and 2.5 eV, correspondingly. For high-quality graphene, the relaxation time at room temperature is about 0.1 ps [80], which corresponds to the energy scale $\Gamma = \hbar/\tau \sim 5$ meV. Graphene chemical potential (or Fermi energy) $\mu_{ch} \approx \hbar v_F(\pi n)^{1/2}$ is determined by surface carrier density $n$ and Fermi velocity $v_F \approx 10^6$ m/s. For example, $n \approx 8*10^{13}$ cm$^{-2}$ corresponds to $\mu_{ch} \approx 1$ eV. We should note that the abovementioned equations assume the electron's dispersion to be linear. This requires that the length of the electron's wavevectors at the Fermi level should be relatively small (typically less than $10^8$ cm$^{-1}$), as compared with the size of the Brillouin zone. This condition is satisfied for small carrier concentration $n \ll 10^{16}$ cm$^{-2}$. Similar estimates show that for graphene's chemical potential of the order of 1 eV the electron's dispersion is also linear and above expressions are applicable.

Moreover, the electromagnetic response of graphene at THz frequencies and the commonly used 0.3-1 eV range of chemical potential is dominated by a simple Drude-like intraband conductivity and the interband contribution can be neglected. In contrast, at the near-infrared and visible frequencies, the interband term plays the crucial role. In the first case, the THz frequency can match the absorption edge, $\hbar\omega_{THz} \approx 2\mu_{ch}$, while in the second (near IR to visible) case we obtain $\sigma_{intra}(\omega) \to 0$, $\sigma_{inter}(\omega) \to e^2/4\hbar$.

Application of the external magnetic field $B$ leads to the charge carriers' circulation in cyclotron orbits and the famous graphene Dirac cone splits in a discrete set of non-equidistant Landau energy levels, see Figure 1A,B. In contrast to the usual semiconductors and metals with parabolic carriers' dispersion, linear dispersion of graphene's carriers leads to non-equidistant Landau levels ($E_j = [2\hbar v^2|eB|j]^{1/2}$, where $j$ is the level number) and include a characteristic zero-energy state ($j = 0$). For magnetooptics, graphene layer has an resonant effects at the photon energies equal to the Landau levels difference. Another feature is associated with the Hall conductivity of graphene, which leads to the Faraday and Kerr effects.

Experimental investigations (see Figure 4) have shown that the Faraday rotation in single layer graphene may reach the giant values of 6 degrees in the magnetic field of 7 Tesla in the far-infrared range [81], about 1 degree in low magnetic fields < 0.7 T in THz frequency range [82], and in magnetic fields < 5 T at microwave frequencies relevant for telecommunication, cell phones, wi-fi etc [83]. Theoretically, conductivity tensor of graphene in external magnetic field was calculated in few works [84, 85], and results based on these theories are in good agreement with the experimental data.

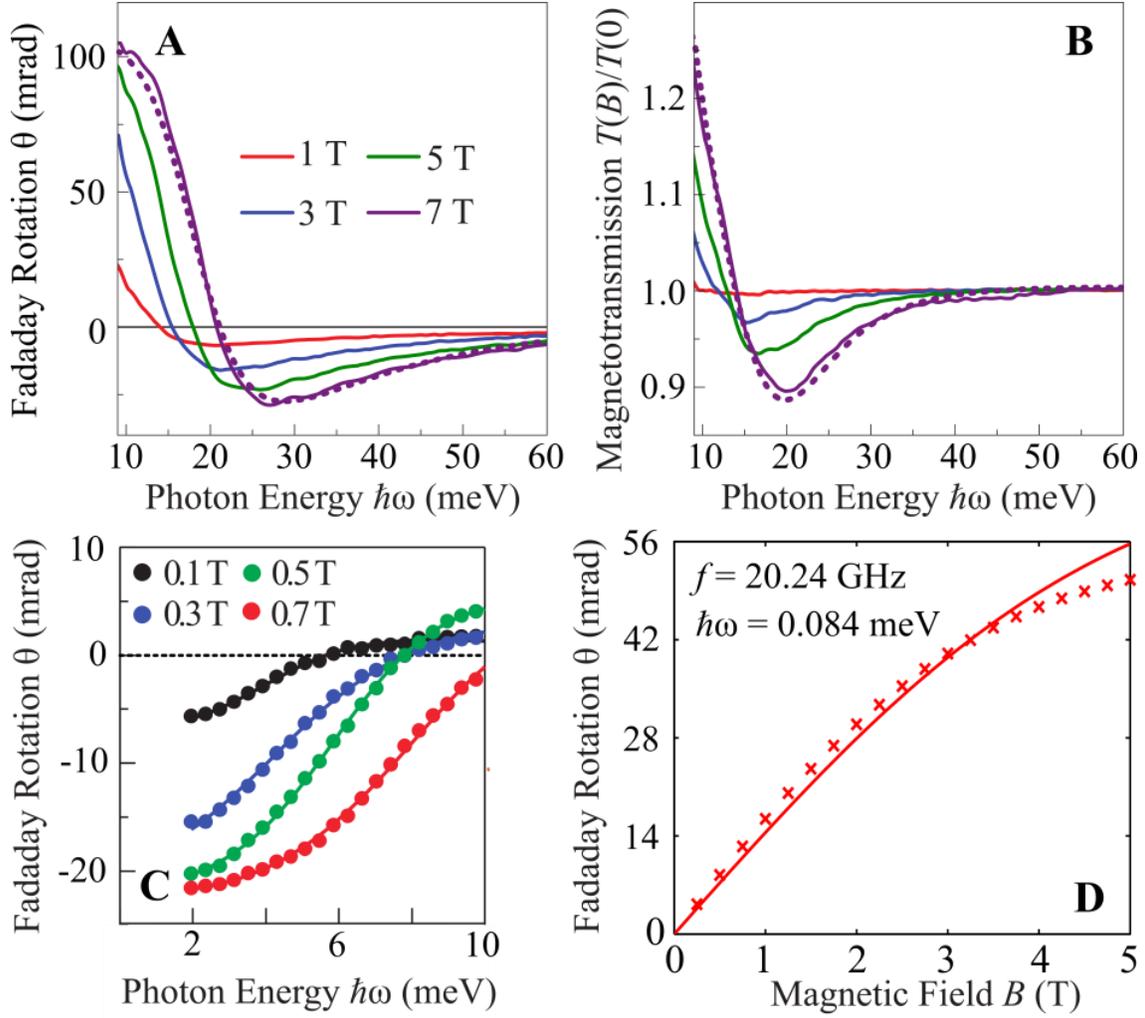

**Figure 4:** Faraday rotation of single layer graphene measured at different frequencies and magnetic fields. (**A, B**) High frequency and high magnetic fields (adapted from [81]), (**C**) lower frequencies and low magnetic fields (adapted from [82]), and (**D**) low frequency and different magnetic fields (adapted from [83]).

The abovementioned peculiarities of the Faraday rotation may be described by the 2D conductivity tensor with non-zero off-diagonal components. In general case, analytical expressions for these components are quite cumbersome, but at low enough frequencies (or high doping level) obeying $\hbar\omega \ll 2\mu_{ch}$, low magnetic fields $E_1 \ll 2\mu_{ch}$, and low temperatures $k_BT \ll 2\mu_{ch}$, they reduce to the classical Drude form

$$\hat{\sigma} = \begin{pmatrix} \sigma & -\sigma_H \\ \sigma_H & \sigma \end{pmatrix},$$

$$\sigma = \sigma_0 \frac{1+i\omega\tau}{(\omega_c\tau)^2 + (1+i\omega\tau)^2}, \quad (3)$$

$$\sigma_H = \sigma_0 \frac{\omega_c\tau}{(\omega_c\tau)^2 + (1+i\omega\tau)^2}.$$

Here, $\sigma_0 = \sigma_{intra}(\omega)$ is given by Eq. (1) and $\omega_c = v_F(2|eB|/\hbar)^{1/2}$ denotes the cyclotron frequency.

To the best of our knowledge, previous investigations of electronic and optical properties in graphene have been restricted to the perpendicular orientation of the magnetic field, which drastically perturbs the trajectories of the electrons moving on their cyclotron orbits. In another case of an in-plane magnetic field the magneto-optical effect in graphene is strongly suppressed for a simple reason. The thickness $d$ of graphene monolayer falls within the range $0.1 - 0.5$ nm [12,14] and the cyclotron radius reads $R_c = mv/eB = \hbar k_F/eB \approx 12(n[10^{10} \text{ cm}^{-2}])^{1/2}/B[\text{T}]$ nm, $\hbar$ is the Plank constant, $n$ is carrier concentration, $k_F$ is Fermi wave vector [42]. For carriers concentrations $n \sim 10^{13}$ cm$^{-2}$ and reasonable values of the magnetic field the cyclotron radius is much larger than graphene thickness: $R_c/d > 10^3/B[\text{T}]$. Therefore, the trajectories of the electrons are almost not perturbed and minor residual magneto-optical effects in graphene do not affect SPPs properties. In a quantum-mechanical picture the quantization of the magnetic flux through such an extremely stretched contour starts playing a role, i.e. the first Landau level emerges also for $B > B_{crit} \sim 10^3$ T.

In a different geometry of a graphene tubes placed in a collinear magnetic field the magneto-optical effects could become important only in a special case when the cyclotron and tube radii become equal. For a tube with the radius $R \sim 200$ nm it would be the case for $B_{crit} \sim 10$ T.

### III. Intrinsic magneto-plasmonics in graphene

Magnetooptical properties of graphene may also give rise to magneto-plasmonics. In contrast to plasmonic excitations which are usually TM-polarized (i.e. the magnetic component of electromagnetic field is perpendicular to the *k*-vector), magneto-plasmons in graphene are hybrid TM-TE modes (all components of electric and magnetic fields are non-zero),

In magnetically biased single-layer graphene, a ensemble of weakly decaying quasi-TE modes, separated by magneto-plasmon-polariton modes, emerges due to the magnetic field [86-88] (see Figure 5A-D). Magneto-plasmons have been experimentally observed in graphene epitaxially grown on SiC [90], in layered graphene structures they even display the drift instability [89].

In a finite width graphene strip, magneto-plasmon-polaritons propagating in the transverse direction may form standing-wave resonances across the strip [91]. Such modes are usually called bulk 2D modes of the graphene stripe. Excitation of these magneto-plasmonic modes may significantly affect magnetooptical response of graphene nano-structures: patterning the uniform graphene layer into graphene stripes array allow to produce the same Faraday rotation at much smaller magnetic fields [92]. The discussed bulk magneto-plasmons and magneto-plasmon-polaritons are intrinsic to extended graphene films where boundary effects are neglected. In finite-size structures edge effects become important as they give rise to the so-called edge magneto-plasmons (localized at the edge) and edge magneto-plasmon-polaritons propagating along the edge [93].

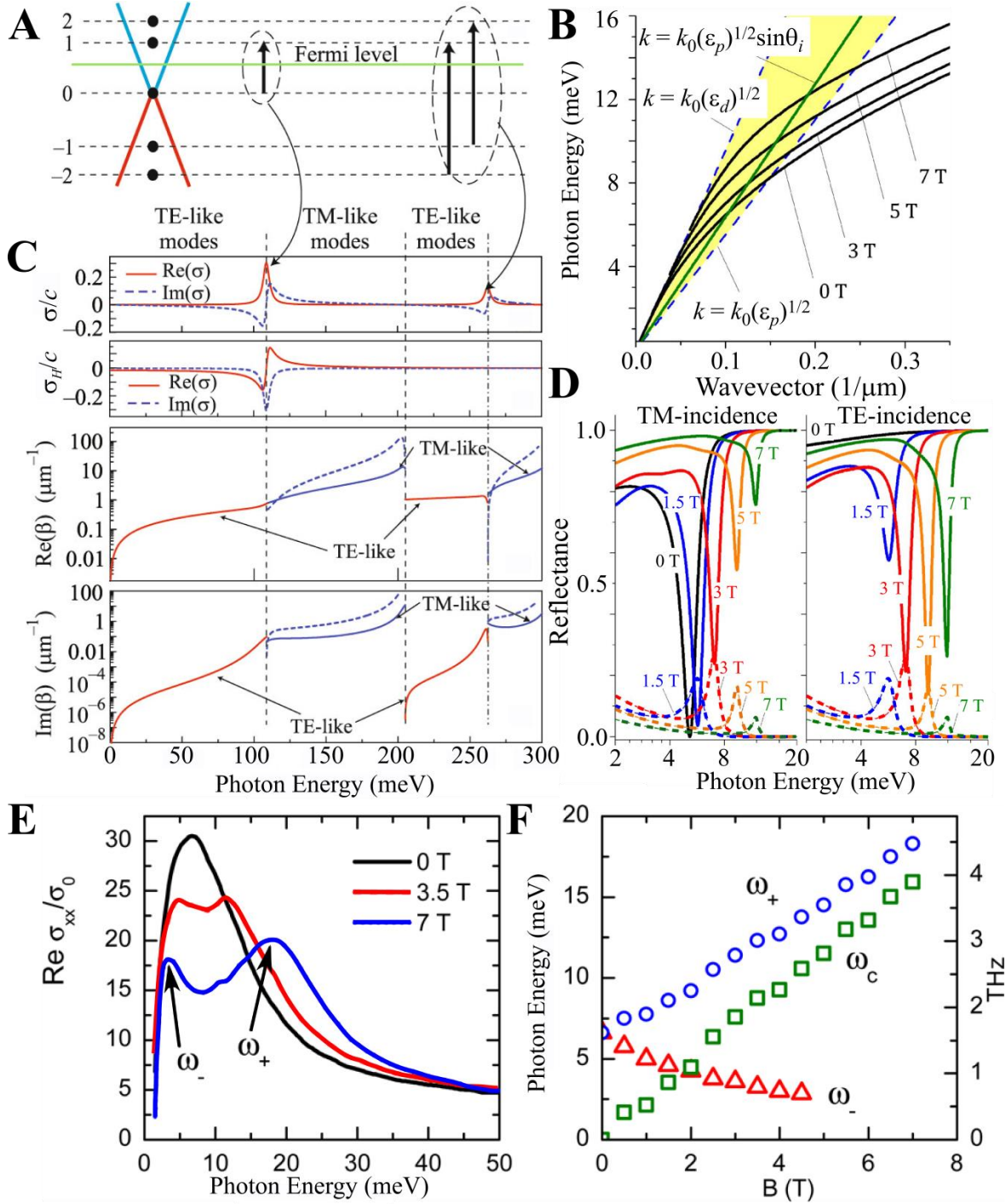

**Figure 5:** Magneto-plasmons in graphene. (**A**) The scheme of Landau levels in graphene. (**B, C**) dispersion curves of the magneto-plasmon-polaritons for different values of an external magnetic field. (**D**) Magneto-plasmon-polaritons in magnetically biased graphene in Otto configuration can be excited by both TE and TM incident wave. In left panel the incident wave is TM polarized, and the solid (dashed) lines correspond to TM (TE) reflectance. In right panel the incident wave is TE polarized, and the solid (dashed) lines correspond to TE (TM) reflectance. (**E, F**) Magnetically biased graphene disks may support bulk ($\omega_+$) and edge ($\omega_-$) magneto-plasmons, (see text for details). (**A, C**) adapted from [87], (**B, D**) adapted from [88], (**E, F**) adapted from [94].

A similar situation is observed in magnetically biased graphene disks [94-96], which support both bulk (localized across the disk) modes and edge (confined to disk edge) one (see Figure 5E, F). Interestingly, these modes are degenerate in magnetic-free graphene disks, while external magnetic field leads to the splitting of edge and bulk magneto-plasmons.

Magnetic field induced breaking of the degeneracy of the edge modes suggests straight-forward applications of the magnetically biased graphene stripes in non-reciprocal plasmonic devices such as phase shifters [91], couplers [97], plasmonic isolators [98], directional SPPs excitation [99], magnetically tunable focusing in planar lenses [100], etc.

## IV. Plasmonics of graphene-based meta-surfaces

As already mentioned before, meta-surfaces are the 2D-analogs of 3D-metamaterials. They consist of subwavelength-sized building blocks (so-called "meta-atoms") periodically arranged on a surface. Such ultrathin structures are promising for light manipulation at the nano-scale: they display an anomalous reflection, diffraction-free propagation, allow to create optical vortexes, manifest the photonic spin Hall effect, etc.

One of the simplest graphene-based meta-surfaces is formed by graphene stripes with the width $A$ separated by the spacer width $B$, with sub-wavelength periodicity $\Lambda = A + B \ll \lambda$ (in electrostatic limit). The optical response of such meta-surface can be described by the highly anisotropic conductivity tensor [35 - 37]:

$$\hat{\sigma}_{meta} = \begin{pmatrix} \sigma_{xx} & \sigma_{xy} \\ \sigma_{yx} & \sigma_{yy} \end{pmatrix}, \qquad (4)$$

where all tensor components depend on graphene conductivity $\sigma_g = \sigma_{intra} + \sigma_{inter}$ with $\sigma_{intra}$ and $\sigma_{inter}$ given by Eqs. (1) and (2) and the capacitive coupling $\sigma_C = i\omega\varepsilon_0\varepsilon_{eff}\Lambda\ln[\sin(\pi B/2\Lambda)]/\pi$ between the stripes. The topological transition from the elliptic to the hyperbolic topology around σ-near-zero case corresponds to $\text{Im}\{A\sigma_C + B\sigma_g\} = 0$. This transition plays a crucial role in plasmonics: whereas in case of elliptic topology, SPPs can propagate in all directions, in hyperbolic case their propagation is allowed is some specific directions only. Physically, the hyperbolic meta-surface displays the metal-like behavior in one direction while showing the dielectric-like response in the orthogonal directions. Due to the pronounced frequency dependence of $\sigma_g$ and $\sigma_C$ the spectral regions of hyperbolic and elliptic topology are separated by a highly anisotropic σ-near-zero point, where large dissipative losses occur.

Figure 6A shows the conductivity components of the meta-surface with the fixed periodicity 50 nm under variation of the stripe width. Hyperbolic regime corresponds to the region with the different signs of the $\sigma_{xx}$ and $\sigma_{yy}$. SPPs excited by a point dipole in the meta-surface working in the hyperbolic regime are shown in Figure 6B.

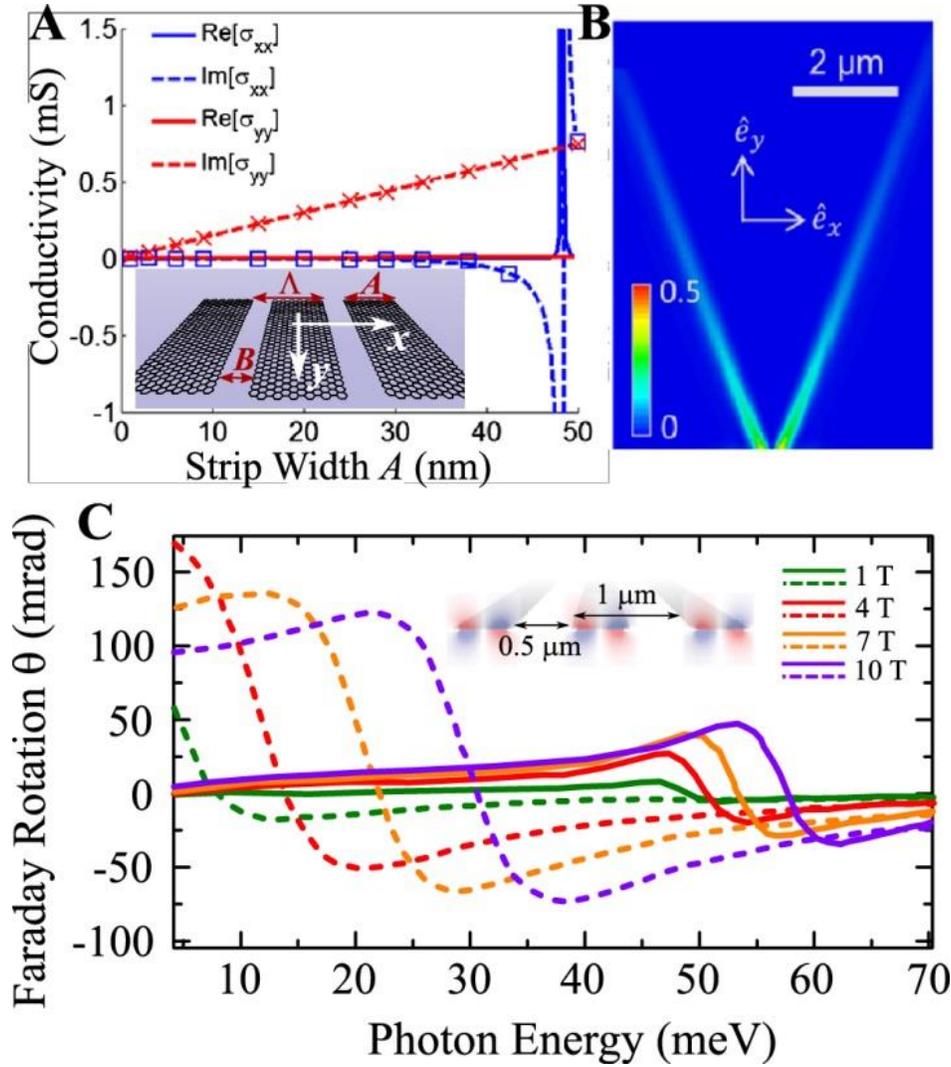

**Figure 6: (A)** Effective conductivity of the meta-surface formed by graphene stripes of the width $A$ placed with the periodicity $W = 50$ nm in long wave limit (i.e. $\lambda_{SPP} \gg W$). Operating frequency is 10 THz. **(B)** SPPs excitation by the point source placed in the meta-surface working in hyperbolic regime. Magneto-plasmon assisted Faraday rotation by a meta-surface placed in a perpendicular external magnetic field **(C)**. Dashed lines correspond to uniform graphene. Figures **(A)** and **(B)** adapted from [36], **(C)** adapted from [92].

As we discussed in the previous Section, excitation of these magneto-plasmonic modes may significantly influence the magnetooptical response of a patterned graphene layer. Figure 6C demonstrates the influence of periodic patterning of graphene and the large magnetic field on the the spectra of Faraday rotation [93]. For an increasing magnetic field the maxima of the dashed (uniform graphene) and solid (patterned graphene) lines in Fig. 6E shift to higher frequencies. The maxima of patterned graphene are also shifted to higher frequencies as compared to the uniform graphene.

V. **Magneto-plasmonics of graphene-based cylindrical structures**

Intrinsic magneto-optical activity of graphene, unfortunately, is useless for cylindrical structures. It needs the static magnetic field directed perpendicular with respect to graphene. In cylindrical geometry, only radial magnetic field will satisfy this condition (or at least magnetic field with non-zero radial component). Such magnetic fields hardly may be obtained in the real experimental conditions. It is worth mentioning that plasmonic structures formed by graphene and magneto-active materials have some interesting features in the planar geometry [101, 102]. So, for introducing magneto-optical activity into cylindrical graphene-based structures one may use magneto-active cylinder covered by graphene. Such a structure has been recently investigated [48] (see Figure 7).

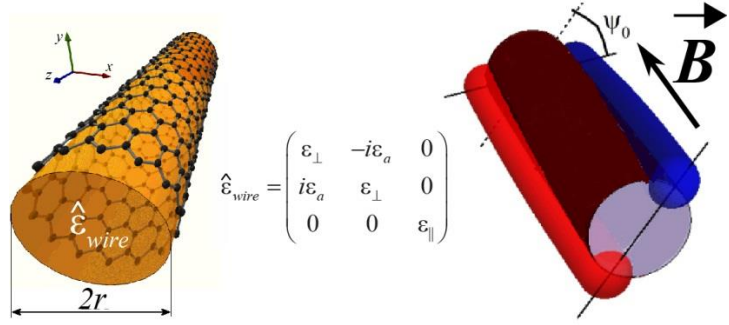

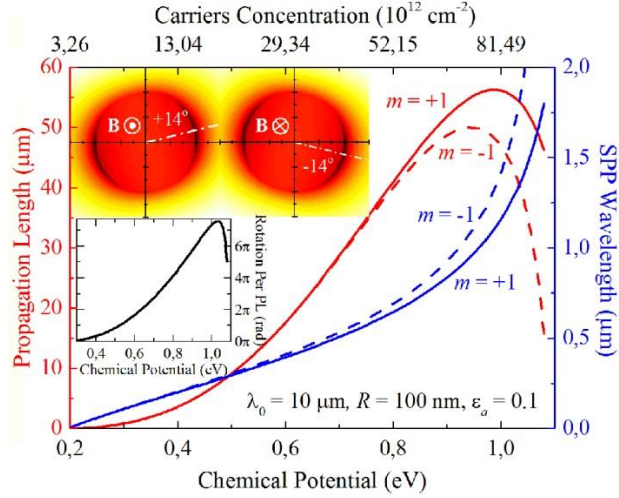

**Figure 7:** A magneto-optical nanowire covered with a graphene layer supports high-order azimuthal SPP modes with $|m| > 0$. Both SPP propagation length (left axis) and wavelength (right axis) for the first-order mode with $|m| = 1$ depend on graphene chemical potential (corresponding carrier concentration is shown on the top axis). Core radius $r = 100$ nm, frequency of electromagnetic wave $f = 30$ THz (vacuum wavelength $\lambda_0 = 10$ μm), permittivity of the core $\varepsilon_\perp = \varepsilon_\parallel = 3$, gyration of the core $\varepsilon_a = 0.1$. The outer medium is vacuum. In the upper inset, the field distributions are shown for the opposite magnetic field directions at $z_0 = 500$ nm and $\mu_{ch} = 1$ eV. Total rotation of the field distribution upon the reversal of the external magnetic field with $\mathbf{B} \approx 1.8$ T (such magnetic field leads to $\varepsilon_a = 0.1$ at Verdet constant $V = 10^4$ rad T$^{-1}$ m$^{-1}$) pointing along the nanowire axis is about 28 degrees. The lower inset shows the figure of merit, i.e. the rotation angle per propagation length. The figure adapted from [48].

SPP modes propagating on the graphene-covered non-gyrotropic nanowires induce the complex distribution of the stationary magnetic field generated via the inverse Faraday effect [100]. At present we assume the SPP intensity is so small that the inverse Faraday effect inside the magnetic nanowire can be neglected. Then the magneto-active core of the cylinder can be described, in the cylindrical coordinate system ($\rho$, $\varphi$, $z$), by the intensity-independent dielectric permittivity tensor with non-zero off-diagonal components $\varepsilon_{\rho\varphi} = -\varepsilon_{\varphi\rho} = i\varepsilon_a$. The cylindrical geometry leads to the azimuthal and axial field distribution of $\mathbf{E_m}$, $\mathbf{H_m} \sim \exp[ih_m z + im\varphi]$, where $h_m = h'_m + ih''_m$ is a complex SPP wavevector (SPP propagation length is $L_{SPP} = 1/h''_m$ and SPP wavelength $\lambda_{SPP} = 2\pi/h'_m$), $m$ is the azimuthal mode index characterizing SPP's chirality. These chiral, azimuthal SPP modes can be understood as plane electromagnetic waves characterized by the longitudinal (along the nanowire axis) and transversal (along the nanowire surface, perpendicular to the nanowire axis) components of the wave vector $k_\parallel = h_{\pm m}$ and $k_\perp = \pm m/r$, respectively. Therefore, the phase front of azimuthal SPP modes is tilted with respect to the nanowire axis by an angle $\alpha_m$ obeying tg $\alpha_m = h_m r/m$. Chiral modes with $|m| > 0$ exist above the cut-off frequency. The number of supported modes at the fixed vacuum wavelength $\lambda_0$ can be estimated as $\mathrm{Re}[i2\pi r(\varepsilon_\perp + \varepsilon_0)c/(\sigma_g \lambda_0)]$. An increase of the core permittivity leads to an increase of the number of supported modes. Whereas for graphene-covered non-gyrotropic nanowires modes with opposite chirality (opposite signs of $m$) are degenerate [16], in the gyrotropic case these modes propagate along the wire at slightly different velocities.

This effect has a simple practical application as a nanoscaled Faraday isolator. Imagine that we excite a gyrotropic graphene-covered nanowire of length $z_0$ by a linearly polarized plane electromagnetic wave at one of the tips. Along with a non-chiral m=0 mode a linear superposition of $\sigma_{+1}$ and $\sigma_{-1}$ modes will be excited [103]. The azimuthal intensity distribution

cos[φ] of this mode superposition will be rotated around the nanowire axis by the angle $\psi = z_0(\text{Re}\{h_-\} - \text{Re}\{h_+\})/2$ and out-coupled into the linearly polarized free-space radiation with the polarization plane rotated by the angle $\psi$. This polarization rotation is characterized by the specific rotation angle $\psi_0 = \psi/z_0$ per unit length. The maximum rotation $\psi_0 L_{SPP}$ is achieved in case of the nanowire length equal to SPP propagation length, $z_0 = L_{SPP}$, and serves as another figure-of-merit in magneto-plasmonics [104].

For the parameters of nanowire radius $r = 100$ nm (quantum effects in graphene structures should be taken into account for the size of the structure less than ≈ 20 nm [105]); $\varepsilon_\perp = \varepsilon_\parallel = 3$, at the lower part of the mid-infrared region ($f = 30$ THz), only one azimuthally dependent mode can be excited. Characteristics of this mode are shown in Figure 7: the total rotation of the field distribution upon the reversal of the external magnetic field pointing along the nanowire axis at $z = 500$ nm and $\mu_{ch} = 1$ eV is about 28 degrees. This value is roughly 30 times larger as compared to the angle of rotation of polarization plane for a plane electromagnetic wave travelling in the volume of the magneto-optical material along magnetization direction $2BVz = 1.04$ deg for $B ≈ 1.8$ T (such magnetic field leads to $\varepsilon_a = 0.1$ at Verdet constant $V = 10^4$ rad T$^{-1}$ m$^{-1}$). Rotation per SPP propagation length $L_{SPP} ≈ 45$ μm may reach a giant value of $8\pi$. The change in sign of the gyrotropy $\varepsilon_a$ (i.e. change in magnetization or magnetic field direction) leads to the opposite rotation of field distribution. Change in graphene conductivity (or its chemical potential) results in the larger difference in propagation constants for modes with opposite signs of $\pm m$. This may be used for adjusting the rotation angle similarly to the graphene-covered optical fiber [47]. Both the specific rotation angle and the rotation angle over SPP propagation length increase with the growing chemical potential. The specific rotation angle is larger for higher-order modes, while the rotation angle over SPP propagation length displays an opposite behavior due to the smaller propagation length of higher-order modes. For maximum rotation angles the propagation lengths of ±*m*-modes differ significantly suggesting that the depth of the azimuthal intensity modulation decreases.

The permittivity of nanowire, its radius and the permittivity of the outer medium may be potentially used for achieving the maximum rotation of the desired mode, but this question has not been investigated in details yet.

### VI. Topological plasmonics of graphene-based meta-structures

As it was pointed above, replacing a continuous graphene layer by a periodic arrangement of graphene strips (graphene meta-surface) may significantly affect the plasmonic properties of the structure. As we are going to show below, the transformation of the meta-surface topology (see Figure 8A for details) from a flat surface into cylinder (meta-tube) and further into a torus (meta-torus) may lead to non-trivial change of plasmonic properties [57].

The transformation of a flat meta-surface in a meta-tube can only be achieved by a finite number of ways (see Figure 8B): the periodic boundary condition with respect to the azimuthal angle and the intrinsic periodicity of the meta-surface give rise to a set of discrete tilt angles

$$\theta_n = \arcsin[n\Lambda/2\pi r]. \qquad (5)$$

An integer number *n*, to be denoted as the "topological index", is the number of graphene stripes winding around the meta-tube. It represents the topological index of the structure because under homeomorphic transformations one cannot change the count of the spirals. The maximum topological index $n_{max} = 2\pi r/\Lambda$ corresponds to the longitudinal orientation of graphene stripes.

Similarly to the previous Section, SPPs propagating along the cylindrical meta-tube are described by electric and magnetic fields $\mathbf{E_m}$, $\mathbf{H_m} \sim \exp[-i\omega t+ihz+im\varphi]$ and modes with opposite ±*m* propagate at different angles with respect to the graphene stripes in our chiral structure.

Calculations show that propagation constants $h_\pm$ for these modes are different as well. We will focus on the modes with $m = \pm1$ (to be denoted as $\sigma_\pm$ due to analogy to the phenomenon of the Faraday rotation in bulk materials) and discuss their dispersion characteristics in detail.

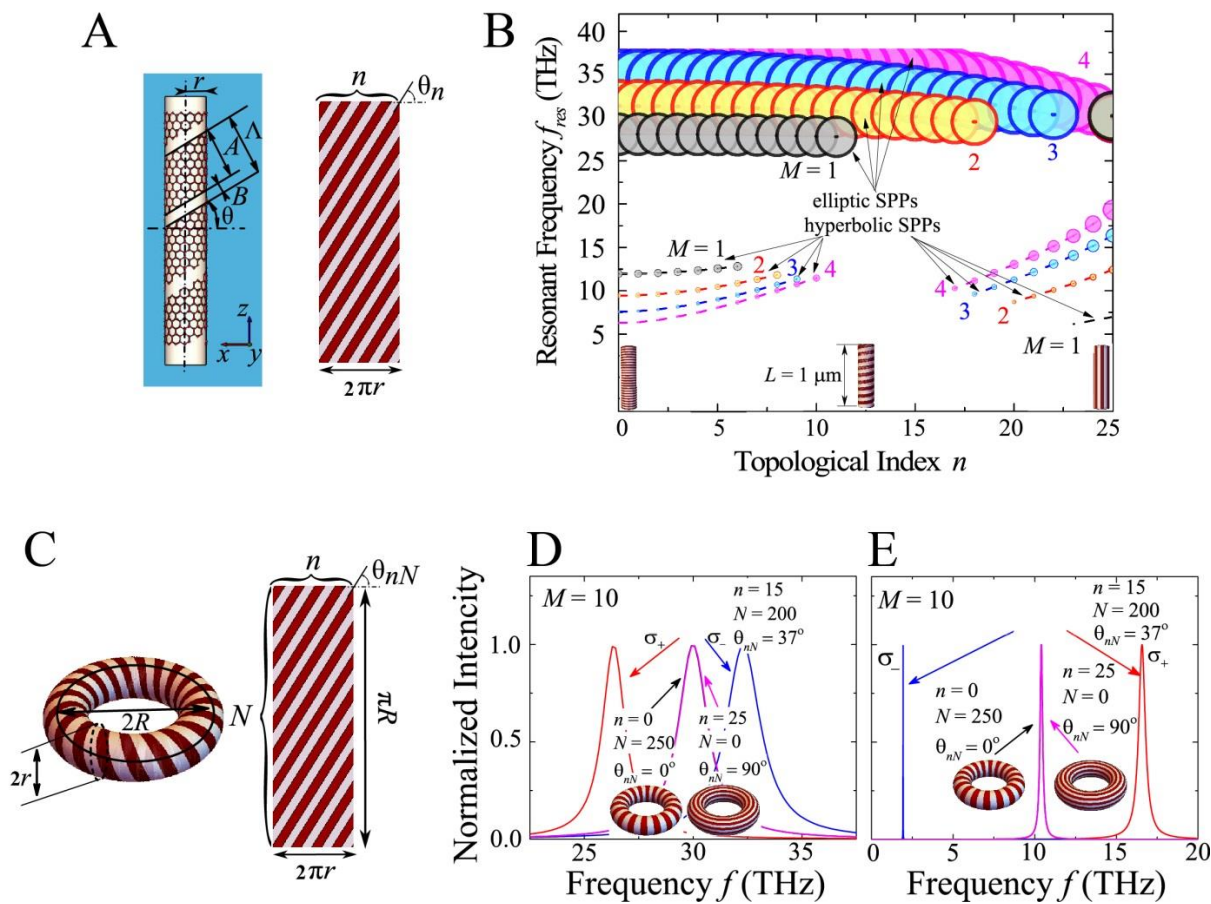

**Figure 8:** Topological transformations of nanostructures formed by meta-surfaces.(**A**). A graphene-based meta-tube (**B**) is obtained by winding a bunch of *n* identical graphene stripes around the cylindrical core under the fixed angle $\theta_n$. A graphene-based meta-torus (**C**) is obtained by winding the meta-tube. It is characterized by pair of indexes *n* and *N*. The resonant frequencies of the finite-length meta-tube (**D**) via the variation of the topological index *n*. Resonance curves (**E**) of the elliptic (left panel) and hyperbolic (right panel) SPPs in the meta-torus correspond to the condition $Rh = M$ for M = 10; $R/r = 10$. The schematic of the torus is shown near each curve. Continuous red and blue lines correspond to the counter-clockwise and clockwise SPP propagation, respectively. Adapted from [57].

In chiral structures the wave vectors of $\sigma_+$ ($m=1$) and $\sigma_-$ ($m = -1$) SPP-modes are oriented differently with respect to graphene stripes. This leads to the difference in their dispersion relation: whereas $\sigma_+$ SPPs possess a cut-off frequency of 10 THz, $\sigma_-$ SPPs are allowed to propagate at lower frequencies. In other words, $\sigma_+$ and $\sigma_-$ dispersion curves are characterized by slightly different bandgaps where SPPs cannot propagate. The bandgap opening is caused by the transition from the elliptic (above the bandgap) to the hyperbolic (below the bandgap) SPPs dispersion. This transition occurs at highly anisotropic σ-near-zero points of the metasurface, where their resonant response is accompanied by large dissipative losses. Obviously, the difference in bandgaps can be used to design structures for the asymmetric, one-way SPP propagation.

In order to justify the importance of $\sigma_\pm$ SPP-modes here we note that a linearly polarized electromagnetic wave impinging on the tip of our meta-tube at $z=0$ will predominantly excite the linear combination of $\sigma_+$ and $\sigma_-$ SPPs with equal amplitudes (i.e. the azimuthal field distribution $\cos[\varphi]$). The field distribution will rotate under propagation similarly to the graphene-covered magnetized nanowire, considered in the previous Section. Adjusting the graphene's chemical potential, though the external gate voltage or intrinsic chemical doping can significantly modify the specific rotation $\psi_0$. The maximum values of $\psi_0$ are expected to be much larger than those predicted in previous Section for the gyrotropic graphene-covered nanowires under similar conditions.

An opposite chirality of the structure can be formally obtained either by assuming negative tilt numbers (or tilt angles) or backward propagating waves.

The non-reciprocal properties of the structure gives the condition for Fabry-Perot resonances in the finite length meta-tube $L[h_+(\omega_{es})+ h_-(\omega_{es})] = 2\pi M$, where $M$ is an integer number. The resonant frequencies are shown in Figure 8B. Both hyperbolic (below the bandgap) and elliptic (above the bandgap) SPPs may form the resonances. Significant shift of the cut-off frequencies for the forward and backward propagating SPPs may prohibit the existence of some lower-order Fabry-Perot modes in chiral meta-tubes because of the one way propagation regime. For the higher-order Fabry-Perot resonances (large $M$), $h$ increases and $\alpha_{\sigma\pm}$ both get close to 90°, when the difference between two polarizations $\sigma_\pm$ ($m=\pm1$) is negligible. This leads to the narrower parameters' range where abovementioned effect may be observed.

In the experiment, a linear combination of fundamental mode ($m=0$) and a chiral mode ($m=+1$ or $m=-1$) can be excited by a linearly polarized light beam focused under an appropriate angle on a tip of a silver nanowire [106-108]. A similar excitation of a chiral meta-tube, with parameters corresponding to the one way propagation regime, should lead to the creation of an electromagnetic "hotspot" at one of the tips [109]; for modes with an opposite chirality the hotspot will be located at the opposite tip.

An even more interesting approach to change the topology of the structure is to transform a cylinder into a torus and thus introduce the second topological index $N$ (see Figure 8C). In such a system, the tilt angle $\theta_{nN}$ must satisfy two distinct conditions: $\theta_{nN} = \arcsin[n\Lambda/2\pi r]$ and $\theta_{nN} = \arccos[N\Lambda/2\pi R]$. For an arbitrary geometry of the torus these conditions are usually not satisfied for an all $n$. For example, two distinct configurations with perpendicular ($n=0$, $N=N_{max}$) and longitudinal ($n=n_{max}$, $N=0$) orientation of graphene stripes exist only for if the ratio $R/r$ of the torus radii is an integer.

The Fabry-Perot condition for modes propagating on a torus clock- and counterclockwise ($2\pi R h_\pm = 2\pi M$) will be satisfied for two different frequencies. Only in two degenerate cases of $0°$ and $90°$ tilt angle the mode propagation would be symmetric and both resonant frequencies would become identical. The resonances of all orders exist in the meta-torus for all possible topological indexes in contrast to the finite length meta-tube. The resonant condition for the meta-torus $Rh = M$ uniquely defines the effective propagation angle of the SPPs: $\tan(\alpha_{mM}) = hr/m = Mr/mR$. In analogy to the topological indices $n$ and $N$ a pair of electrodynamic indices $m$ and $M$ defines the electrodynamic topology of the resonant mode. An existence of topological SPP resonances on a meta-torus implies a fixed relation between the structural topological indices ($n$, $N$) and electromagnetic topological indices ($m$, $M$) of a resonant mode:

$$\frac{n}{N}\operatorname{ctg}\theta_{nN} = \frac{m}{M}\operatorname{tg}\alpha_{mM} \qquad (6)$$

In general, elliptical and hyperbolic $\sigma_\pm$ modes possess different frequencies for all structures. An exception is an exotic case when SPP's propagation angle $\alpha_{\sigma\pm} = 45°$ with respect to graphene stripes is identical for longitudinal and perpendicular stripe orientation. In this case, the splitting of resonant elliptic and hyperbolic $\sigma_\pm$ modes for a chiral structure is almost the largest one, and Eq. (6) helps to explain such behavior. The analysis of SPPs propagating on a flat metasurface shows that their frequencies display the largest difference for wavevectors along and perpendicular with respect to graphene stripes, respectively. For resonant modes in the chiral toroidal resonator we have equal $h_{\sigma\pm}$, and thus, $\alpha_{\sigma-} = 180° - \alpha_{\sigma+}$. In analogy to planar metasurfaces, the maximum frequency splitting in chiral toroidal structure should be achieved for $\alpha_{\sigma-} - \alpha_{\sigma+} = 90°$, which leads to $\alpha_{\sigma+} = 45°$, $\alpha_{\sigma-} = 135°$; in addition we should have $\theta_{nN} = 45°$. Under these conditions we can use Eq. (6) to calculate the set of topological indices for the maximum frequency splitting obeying $Mn/N = 1$. The resonant curves in the vicinity to this condition are shown in Figure 8E.

## VII. Concluding remarks

We have reviewed the most recent results in plasmonics of magnetic and topological graphene-based nanostructures. The giant Faraday rotation of high-order plasmonic modes in graphene-covered nanowire and their tuning by both the gyrotropy (magnetic field or magnetization) and graphene chemical potential (chemical doping or gate voltage) may be used to magnetically control the density of states of electromagnetic radiation at the deeply sub-wavelength length scale, an effect interesting for quantum-optical devices operating in the Telecom frequency range [110]. In nanowires of finite length the magnetic field can be exploited to tune the Fabry-Perot cavity modes, an effect going beyond the resonant enhancement of Faraday rotation in the magneto-optical medium positioned inside an optical resonator [111].

While a graphene meta-tube displays a similar rotation of azimuthal plasmonic modes, the larger magnitude of the rotation per unit length plays a crucial role for the design of "one-way propagation" plasmonic devices: it is responsible for the disappearance of Fabry-Perot resonances in finite-length meta-tubes. A piece of a meta-tube, rolled in a torus, possesses a distinct spectrum of azimuthal cavity modes with an unexpectedly large energy splitting between the clock- and counterclockwise propagating SPP modes. Interestingly, the electromagnetic and geometrical topological indices of the structure are intimately connected by simple analytical expressions, which physical meaning remains to be clarified.

In this review we have discussed two kinds of topology. The first one is the topology of the isofrequency contours (at one frequency) observed on hyperbolic graphene meta-surfaces. The second one is a geometrical topology that characterizes the geometry of meta-cylinders and meta-tori. It is characterized by an integer topological index $n$, which determines the electromagnetic behavior of plasmons.

The third kind of topology, which we have not discussed in this review article, is related to the topology of the bands (several frequencies) over the entire Brillouin zone. It is quantified by the so-called Chern number, which change implies a topological phase transition that corresponds to drastic change in the physical properties of the system (see Ref. 112). Recent investigations of the topological aspects of the third kind (i.e. Chern numbers, Berry phase, Berry connection etc.) of electromagnetic waves propagation in bianisotropic continuous media attracted considerable interest as well [113, 114]. For example, as the bulk transparent magneto-optical material is continuously transformed in isotropic opaque plasma, the evolution of band structure Chern numbers of individual electromagnetic modes experience the jumps [114]. In another relevant example of SPPs propagating along the interface between an isotropic plasma and anisotropic (magneto-optical) dielectric material, one-way propagation of SPPs is found [114], whereas the propagation direction is uniquely determined by the direction of an external magnetic field. Such systems with broken time-reversal symmetry are sometimes designated as Chern-type insulators (the analogs of quantum Hall insulators). Some recent works have been devoted to the investigation of topologically protected (plasmonic) edge states in 2D materials and particularly graphene with the magnetically broken time reversal symmetry [115, 116].

It is worth mentioning that the interplay between the chirality and the hyperbolicity in bulk metamaterials may significantly influence on the abovementioned topology of the third kind and protect the edge states from backscattering even without the requirement to break the time-reversal symmetry [117].

The similarity of the effects for magnetized and chiral graphene-based cylinders raises a question of the possibility of magnetic compensation of the chirality effect in hybrid magneto-chiral plasmonic structures (see Figure 8). The interplay between chirality and gyrotropy in magnetized chiral waveguides formed by plasmonic nano-ellipsoids has been shown to strongly enhance the nonreciprocal response of the structure [118]. In the cylindrical geometry proposed here we expect a non-trivial interference between the geometrical chirality and gyrotropy on SPPs with opposite electromagnetic chirality $m$ or propagating in opposite directions. It might manifest itself in a different dependence of $h_+(B)$ and $h_-(B)$ on the amplitude $B$ of the external magnetic field. The broken time-reversal symmetry can be achieved not only by magnetic effects discussed above but also through the optical pumping [119, 120].

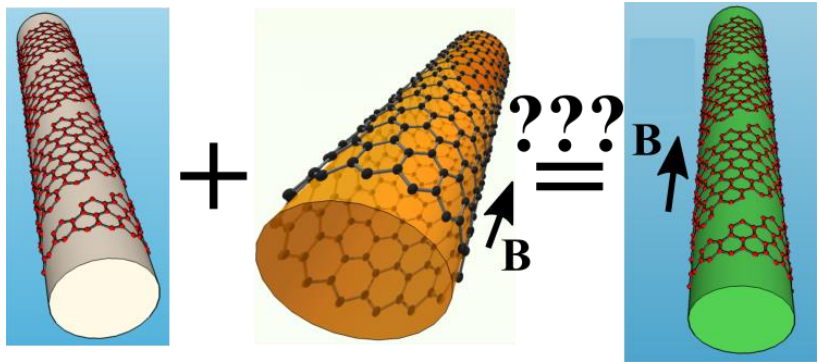

**Figure 8:** The concept of hybrid magneto-chiral nanostructures for possible non-reciprocal magnetic compensation of the chirality effects.

Another open problem is the self-consistent propagation of high-intensity SPPs in systems with magneto-optical materials. Whereas the magnetization naturally affects the SPPs characteristics via the (linear) magneto-optical effect, the SPPs may also affect the magnetization via inverse (nonlinear) magneto-optical effects. The inverse Faraday effect generates the magnetic field $\delta B(E) \sim E^2$ and introduces the nonlinear correction to the off-diagonal permittivity components

$$\varepsilon_a \rightarrow \varepsilon_a + \beta E^2, \tag{7}$$

an effect similar to the third order non-linearity is expected in Maxwell's equations [121]. The complexity of field distributions in cylindrical systems [103] makes it difficult to solve this problem analytically. However, this non-trivial possibility of combining transient magnetic effects with the nonlinear optics, without using an external magnetic field, represents an interesting problem for computational electrodynamics.

An interesting perspective provided by the possibility to apply external strains to meta-structures. Let's consider a non-chiral structure formed by a nanowire longitudinally covered by graphene stripes, where no rotation of high-order plasmonic modes can be observed. If shear strains are applied to such a structure, the spiral waveguide will be formed with the tilt angle defined by the strain value. Alternatively, an axial strain applied to the chiral structure will lead to the change in the spacer width and the periodicity of the structure, or equivalently to change of the filling factor. In any case, the rotation angle of the electromagnetic field distribution at the output of the meta-tube will be modified by the applied strain. For a finite-length meta-tube frequencies of Fabry-Perot resonances frequency will be shifted or even some resonances may vanish. These effects may be useful for future applications as plasmonic strain sensors.

An even more intriguing effect of elastic deformations on plasmonic properties may be envisaged in spatially uniform graphene-based nanowires. We have pointed above that in cylindrical geometry the intrinsic magneto-plasmonic properties of graphene cannot be used, because the radial magnetization is needed. However, it is well known that external strains may affect electronic properties of graphene similarly to an external magnetic field (i.e. strains may induce the so-called pseudo-magnetic field) [122]. Thus, the investigation of acousto-magneto-plasmonic effects in graphene-based nanowires extends beyond the straight-forward effects dominated by the influence of core magnetization on SPPs propagating along graphene nanowires.

**Acknowledgements**

The work was financially supported in part by RFBR (16-37-00023, 16-07-00751, 16-29-14045, 17-57-150001), RScF (14-22-00279), Grant of the President of the RF (МК-1653.2017.2), Act 211 Government of the Russian Federation (contract № 02.A03.21.0011), Stratégie internationale NNN-Telecom de la Région Pays de La Loire, PRC CNRS-RFBR "Acousto-magneto-plasmonics" and the PhOM Research Seminars Program of the Université Paris-Saclay.

AUTHOR INFORMATION

**Corresponding Authors**

*E-mail: kuzminda@csu.ru, vasily.temnov@univ-lemans.fr